\documentclass[11pt]{article}
\usepackage{amsmath,epsf,epsfig,float,amssymb,latexsym}
\usepackage{graphics,psfrag,cite}
\newcommand{\cO}{{\cal O}}

\newcommand{\vp}{{\mathbf{p}}}

\newcommand{\vk}{{\mathbf{k}}}

\newcommand{\bg}{\begin{align}}
\newcommand{\eeg}{\end{align}}
\newcommand{\be}{\begin{equation}}
\newcommand{\ee}{\end{equation}}
\newcommand{\ba}{\begin{eqnarray}}
\newcommand{\ea}{\end{eqnarray}}

\newcommand{\nn}{\nonumber}

\newcommand{\vs}{\vspace{-0.2cm}}
\newcommand{\la}{\langle}
\newcommand{\ra}{\rangle}

\numberwithin{equation}{section}

\begin{document}

\thispagestyle{empty}
\vspace{0.5cm}
\hfill{\tiny HISKP-TH-10/15, FZJ-IKP(TH)-2010-13}
\vspace{2cm}

\begin{center}
{\Large{\bf The chiral quark condensate and pion decay constant\\[0.3em]in nuclear matter at next-to-leading order}}
\end{center}
\vspace{.5cm}

\begin{center}
{\Large A. Lacour$^{a}$, J. A. Oller$^{b}$ and U.-G. Mei{\ss}ner$^{a,c}$}
\vskip 10pt
{\it  $^a$Helmholtz-Institut f\"ur Strahlen- und Kernphysik (Theorie) and
  Bethe Center for Theoretical Physics\\ Universit\"at Bonn,
D-53115 Bonn, Germany}\\
{\it  $^b$Departamento de F\'{\i}sica, Universidad de Murcia, E-30071 Murcia, 
Spain}\\
{\it $^c$Institut f\"ur Kernphysik, Institute for Advanced Simulation and
J\"ulich Center for Hadron Physics\\Forschungszentrum J\"ulich, D-52425 J\"ulich,
Germany}
\end{center}

\vspace{1cm}
\noindent
\begin{abstract}
Making use of the recently developed chiral power counting for the physics 
of nuclear matter \cite{nlou,Lacour:2009ej},
we evaluate the in-medium chiral quark condensate up to next-to-leading order
for both symmetric nuclear matter and neutron matter. Our calculation includes 
the full in-medium  iteration of  the leading order local and one-pion
exchange nucleon-nucleon interactions. Interestingly, we find a cancellation 
between the contributions stemming from the quark mass dependence of the
nucleon mass appearing in the in-medium nucleon-nucleon interactions. Only 
the contributions originating from the explicit quark mass dependence of the pion mass 
survive. This cancellation is the reason of previous
observations  concerning the dominant role of the long-range pion
contributions and the suppression of short-range nucleon-nucleon
interactions. We find that the linear density contribution to the 
in-medium chiral quark condensate is only slightly modified for pure neutron 
matter by the nucleon-nucleon interactions. For symmetric nuclear matter the 
in-medium corrections are larger, although smaller compared to other
approaches due to the full iteration of the lowest order nucleon-nucleon 
tree-level amplitudes. Our calculation satisfies the Hellmann-Feynman 
theorem to the order worked out.
Also we address the problem of calculating the leading in-medium corrections to the pion decay constant.
We find that there are no extra in-medium corrections that violate the   Gell-Mann--Oakes--Renner relation up to next-to-leading order.
\end{abstract} 

\newpage

\section{Introduction}
\label{sec:im_cqc.intro}

The QCD ground state, or vacuum, is characterized by the presence of a strong condensate $\langle  0|\bar{q}q|0\rangle$ of scalar quark-antiquark pairs -- the \textit{chiral quark condensate} -- which represents an order parameter for spontaneous chiral symmetry breaking in QCD.
It is accepted on phenomenological grounds that the lightest pseudoscalar mesons, the pions, are identified with the Goldstone bosons of the spontaneously broken chiral symmetry \cite{nambu}.
Spontaneous chiral symmetry breaking results because the axial charge, $Q_A^i=\int d^3x A_0^i(x)$, does not annihilate the ground state. 
The coupling of the Goldstone bosons to the axial-vector charge is given in terms of the pion decay constant $f_\pi$, which determines the chiral scale $4\pi f_\pi \sim 1$~GeV that governs the size of the quantum corrections in Chiral Perturbation Theory ($\chi$PT) \cite{georgi}. 
Chiral symmetry is also explicitly broken due to the small masses of the lightest quarks, $u$ and $d$.
An interesting issue for the discussion of the QCD phase diagram is the dependence of the chiral quark condensate on temperature and density.
It is expected that with growing temperature, the quark condensate melts \cite{tem1,tem2,Bernard:1987ir}.
There is an indication that this is also the case at zero temperature and increasing density as follows from the direct application of the Hellmann-Feynman theorem \cite{hell1} to the energy density of a 
Fermi sea.
A restoration of chiral symmetry is believed to be linked to a phase transition in QCD.
A nonvanishing chiral quark condensate represents a sufficient condition for spontaneous breakdown of chiral symmetry.
Note, that it is not a necessary one, such one would be the Goldstone boson (pion) decay constant, therefore chiral symmetry might as well be broken with a vanishing quark condensate.
As long as the matrix element $\la \Omega|Q_A^i|\pi^a(\vp)\ra=i\delta_{ia}(2\pi)^3 f_t p_0 \delta(\vp)$ is not zero, the ground state is not left invariant by the action of the axial charge $Q_A^i$ and spontaneous chiral symmetry breaking happens.
In the previous equation $|\pi^a(\vp)\ra$ denotes a pion state with Cartesian coordinate $a$, three-momentum $\vp$, energy $p_0$ and $f_t$ is the temporal weak pion decay coupling.
Many recent calculations in nuclear matter share the assumption that spontaneous chiral symmetry breaking still holds for finite density nuclear systems.
The form of the chiral Lagrangians changes depending whether the chiral quark condensate $\la \Omega| \bar{u}u+\bar{d}d|\Omega\ra$ is large or small.
In the former case we have standard $\chi$PT \cite{wein,gl1,gl2} and in the latter generalized $\chi$PT \cite{jan1,jan2} should be employed.
It has been shown that the first case holds \cite{prlbern} in vacuum for SU(2) $\chi$PT.
For modern applications of chiral symmetry to nuclear systems we refer to \cite{Epelbaum:2008ga}.
The in-medium chiral quark condensate for symmetric nuclear matter in the linear approximation \cite{hell2,Thorsson:1995rj,Wirzba:1995sh,Kirchbach:1996xy,wirzba,annp}   decreases  with density as $1-(0.35\pm 0.09)\rho/\rho_0$, with the error governed by that of the 
knowledge in the pion-nucleon sigma term, $\sigma=\hat{m} \partial m/\partial \hat{m}=45\pm 8~$MeV \cite{Gasser:1990ap,Koch:1982pu}, with $\hat{m}=(m_u+m_d)/2$ the mean of the $u$ and $d$ quark masses, $m$ the nucleon mass and $\rho_0\simeq 0.16$~fm$^{-3}$ the nuclear matter saturation density.
Thus, if the low energy linear decrease in the quark condensate is extrapolated to higher densities the quark condensate would vanish for $\rho=(2.9\pm 0.7)\rho_0$ and  standard $\chi$PT would not be appropriate. 
Of course, higher order corrections could spoil this tendency.
Hence, the calculation of the in-medium quark condensate is  of great interest beyond the linear approximation.
The quark condensate in the nuclear medium and the quest for restoration of chiral symmetry at finite baryon density (and temperature) has a long and outstanding history, see e.g.\ \cite{Hatsuda:1985ey,Reinhardt:1987da,hell2,Thorsson:1995rj,Wirzba:1995sh,Brockmann:1996iv,Kirchbach:1996xy,wirzba,lutz,Huang:2000cr,annp,Tsushima:2006yc,Kaiser:2007nv,Kaiser:2008qu,tubinguen}. 

In this work we concentrate 
 first 
on the application of in-medium baryon $\chi$PT to the calculation of the quark condensate and go beyond the linear density approximation.  
However, our approach \cite{nlou} compared with previous works offers two novel features: i) It follows a strict chiral power counting that takes into account both long- and short-range (multi-)nucleon interactions. It is applicable both in vacuum and in the medium so that a clear connection between the two cases is established and used. ii) We do not take as starting point the Hellmann-Feynman theorem but directly apply the power counting mentioned to the problem of the calculation of the quark condensate. The fulfilment of the Hellmann-Feynman theorem is a consequence of the consistency of the approach order by order. In this way, the dependence on the quark mass of the input parameters in the theory, like $f_\pi$, $g_A$, nucleon and pion masses, etc, is built in. This power counting has been previously applied to the problems of the pion self-energy in the nuclear medium and the equation of state of nuclear matter. For the former issue \cite{nlou} we found a cancellation of  the next-to-leading order (NLO) contributions, including those stemming from the in-medium nucleon-nucleon interactions. This suppression is interesting since it allows to understand from first principles the phenomenological success of fitting data on pionic atoms with only meson-baryon interactions \cite{weiprl,galrep,ericeric}. Regarding the equation of state of nuclear matter it is interesting to point out that saturation for symmetric nuclear matter and repulsion for neutron matter is obtained in ref.~\cite{Lacour:2009ej} at NLO. The resulting curves of the energy per particle as a function of density for neutron matter agrees with sophisticated many-body calculations \cite{urbana} in terms of a subtraction constant $g_0$ that acquires its natural expected value, close to $-0.5~ m_\pi^2$. For the case of symmetric nuclear matter the experimental properties of the minimum for the bounding energy are reproduced fitting  $g_0\simeq -m_\pi^2$, while the analogous parameter $\widetilde{g}_0$ keeps its just mentioned natural value. 
As a result of the present and the previous applications \cite{nlou,Lacour:2009ej,annp} the calculation of the in-medium 
corrections to the coupling of the pion to the axial-vector current up to NLO is 
straightforward.  As we show below, the linear reduction with density of $f_t$ as obtained in 
ref.~\cite{annp} holds up to NLO.

After this introduction,
 we present the power counting formula  used and the Feynman diagrams contributing to the in-medium chiral quark condensate to NLO in section~\ref{sec:pc}.
The contributions stemming from pion-nucleon chiral dynamics are the object of section~\ref{sec:im_cqc.pc}. The terms due to the 
 nucleon-nucleon interactions are calculated  in section~\ref{sec:im_cqc.vrho2}.
Results and discussions thereof are given in section~\ref{sec:im_cqc.num}. 
The in-medium pion decay constant and the Gell-Mann--Oakes--Renner relation
are discussed in section~\ref{sec:im_pdc.disc}. The last section is dedicated to
 offer our conclusions.

\section{In-medium chiral power counting and diagrams}
\label{sec:pc}

For determining the set of diagrams to be calculated in the evaluation of the quark 
condensate in the nuclear medium up-to-and-including NLO contributions we make use of the 
chiral power counting developed in \cite{nlou}
\begin{align}
\nu&=4-E+\sum_{i=1}^{V_\pi}(\ell_i+n_i-4)+\sum_{i=1}^V(d_i+v_i+\omega_i-2)+V_\rho~.
\label{fff}
\end{align}
In this way, $E$ is the total number of external pion lines and $V_\pi$, $V$ and $V_\rho$ are the meson-meson, 
meson-baryon and in-medium generalized vertices \cite{prcoller}, in this order.  In simple terms, an in-medium
 generalized vertex corresponds to a closed nucleon loop that could contain an arbitrary number of meson-baryon
 vertices inserted. In the first summation on the right-hand-side of the previous equation the symbols  $\ell_i$ and $n_i$ are the 
chiral order and number of pionic lines of  the $i^{th}$ meson-meson vertex, respectively. In the sum over the 
meson-baryon vertices, $d_i$ is the chiral dimension of the $i^{th}$ meson-baryon vertex,
 $v_i$ is the total number of mesons lines attached to it, including both pion and heavy meson lines, 
while $\omega_i$ is the number of the  latter ones only. The heavy meson lines correspond to  
auxiliary fields responsible for the local multi-nucleon interactions when taking their masses to infinity \cite{nlou,Borasoy:2006qn}.

Eq.~\eqref{fff} counts every nucleon propagator as ${\cal O}(p^{-2})$ instead of ${\cal O}(p^{-1})$, as
 in the standard counting used in Baryon $\chi$PT \cite{sainio,intj}. In this way, the infrared enhancements 
associated to the large nucleon mass are taken into account from the onset. Despite baryon 
propagators are counted as ${\cal O}(p^{-2})$ it is important to stress that eq.~\eqref{fff} is bounded 
from below \cite{Lacour:2009ej}. According to eq.~\eqref{fff}   the number of lines in a diagram
 can be augmented without increasing the chiral power \cite{nlou} by adding i) pionic lines attached
 to lowest order mesonic vertices, $\ell_i=n_i=2$, ii) pionic lines attached to lowest order meson-baryon
vertices, $d_i=v_i=1$ and iii) heavy mesonic lines attached to lowest order 
bilinear vertices, $d_i=0$, $\omega_i=1$. In this way, resummations of infinite strings of diagrams are required, 
and these resummations signal the appearance of non-perturbative physics. As a major difference between 
 this counting and the Weinberg 
one \cite{wein1,wein2}, the former applies directly to the physical amplitudes while the latter does only to the potential. 
It is also important to stress that $\nu$ in eq.~\eqref{fff} is a lower bound for the actual chiral power of
 a diagram, $\mu$, with $\mu\geq \nu$. The chiral order might be higher than $\nu$ because 
nucleon propagators, always counted as ${\cal O}(p^{-2})$ in eq.~\eqref{fff}, could actually count 
as ${\cal O}(p^ {-1})$ in some cases. If this is so, every ${\cal O}(p^{-1})$ baryon propagator 
 increases the actual chiral order by one unit as compared with $\nu$ eq.~\eqref{fff}. The important point is that all the diagrams with a chiral 
order $\mu$ are a subset of the diagrams with $\nu\leq \mu$.  We refer to \cite{nlou} for more details on eq.~\eqref{fff},  including its derivation.

\begin{figure}[t]
\psfrag{Vr=1}{{\small $V_\rho=1$}}
\psfrag{Vr=2}{{\small $V_\rho=2$}}
\psfrag{Op5}{{\small $ {\cal O}(p^5)$}}
\psfrag{Op6}{{\small $ {\cal O}(p^6)$}}
\psfrag{i}{{\small $i$}}
\psfrag{j}{{\small $j$}}
\psfrag{q}{{\small $q$}}
\centerline{\fbox{\epsfig{file=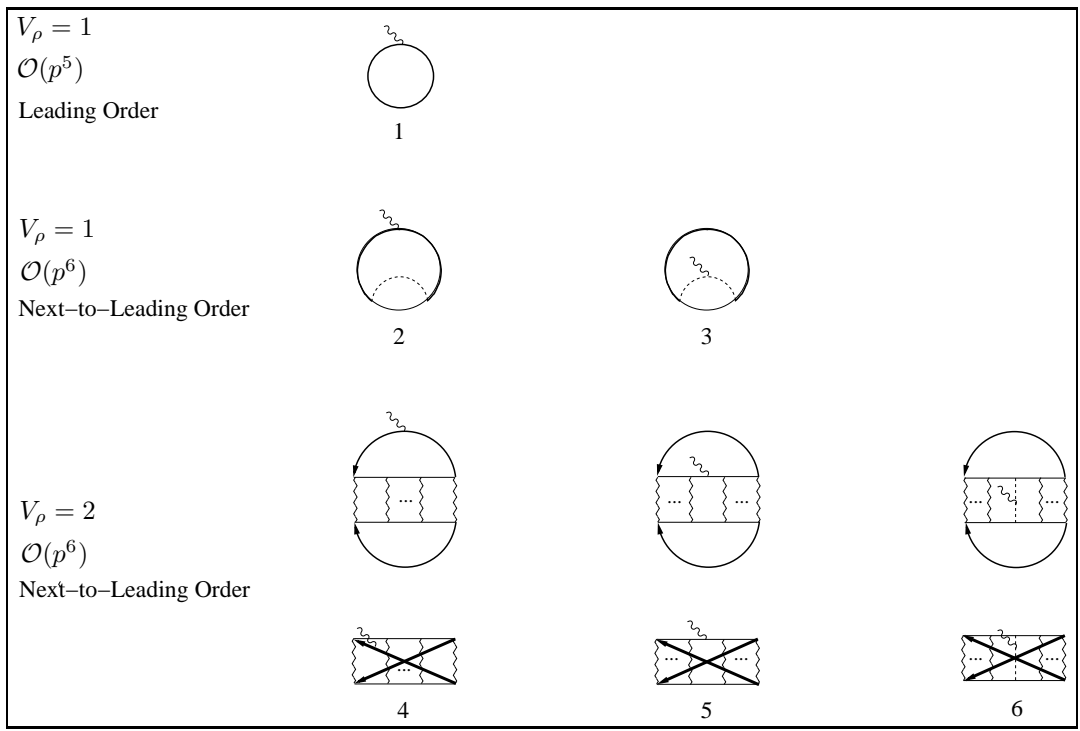,width=.8\textwidth,angle=0}}}
\vspace{0.2cm}
\caption[pilf]{\protect \small
Contributions to the in-medium chiral quark condensate up to  NLO or ${\cal O}(p^6)$. The scalar source with zero 
momentum is indicated by the wavy line and pions by the dashed ones. A wiggly line corresponds 
to the nucleon-nucleon interaction kernel (given in fig.~\ref{fig:wig}) whose iteration is denoted by the ellipsis. 
\label{fig:all_imcqc}}
\end{figure} 

The explicit chiral symmetry breaking due to the quark masses is incorporated
by $s(x)={\cal M}+\delta s(x)$ with ${\cal M}={\rm diag}(m_u,m_d)$ the quark mass matrix.
From the generating functional in the presence of external sources, ${\cal Z}(v,a,s,p)$, the quark condensate is obtained by partial functional differentiation
\be
 \la \Omega | \bar{q}_i q_j | \Omega \rangle = - \frac{\delta}{\delta s_{ij}(x)}{\cal Z}(v,a,s,p)|_{v,a,s,p=0} ~,
 \label{g.qc}
\ee
with $q_1=u$ and $q_2=d$ quarks.
Notice that the quark condensate has a global minus sign compared to diagrams calculated by using ordinary Feynman rules.
See ref.~\cite{prcoller} for a derivation of the generating 
functional ${\cal Z}(v,a,s,p)$   in the nuclear medium making use of functional methods, although keeping only pion-nucleon interactions. 
We also refer to  \cite{annp} for more details on the use of external sources in relation with in-medium $\chi$PT calculations. 

The quantity we will calculate in the following is the in-medium correction $\Xi$ to the chiral quark condensate given by
\begin{equation}
  m_q \langle \Omega|\bar{q}_iq_j|\Omega\rangle  = m_q \langle 0|\bar{q}_i q_j|0\rangle + m_q \, \Xi ~,
\label{xi}
\end{equation}
where $m_q$ is the mass of a certain quark flavor.
Furthermore, the nuclear matter ground state is denoted by $|\Omega\ra$, to distinguish it from 
the vacuum ground state $|0\ra$. 
The resulting contributions are denoted by $\Xi_i$ and are shown in fig.~\ref{fig:all_imcqc}.
To determine the set of diagrams needed for the calculation of the in-medium chiral quark condensate up to NLO 
we proceed by increasing $V_\rho$ step by step in eq.~\eqref{fff}. For each $V_\rho$ we then determine the possible configurations 
of vertices and lines according to eq.~\eqref{fff}. The resulting diagrams are shown in fig.~\ref{fig:all_imcqc}. They can 
easily be identified by considering the corresponding vacuum diagrams. For $V_\rho=1$ the first diagram is the lowest-order 
contribution to the nucleon sigma-term. The rest of diagrams in fig.~\ref{fig:all_imcqc} are NLO. The diagrams 2 and 3 stem from the 
one-pion loop self-energy with the scalar source attached to a nucleon or pion propagator, respectively. 
For $V_\rho=2$ one can conveniently think of the nucleon-nucleon scattering in the presence of a scalar source in vacuum. 
Then proceed by closing the diagrams, which corresponds to sum over the states in the Fermi seas of the nucleons.
The upper row of diagrams~4, 5 and 6 in fig.~\ref{fig:all_imcqc} involves the direct nucleon-nucleon interactions
while the lower row originates from the exchange part. In the following we indicate by $\Xi_i$ the 
contribution to the in-medium quark condensate due to the diagram~$i$ in fig.~\ref{fig:all_imcqc}.
\begin{figure}[ht]
\centerline{\epsfig{file=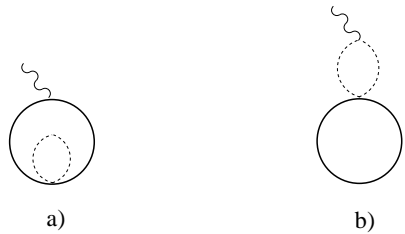,width=.3\textwidth,angle=0}}
\vspace{0.2cm}
\caption[pilf]{\protect \small
These diagrams are zero because of the antisymmetric structure of the Weinberg-Tomozawa vertex coupling in the pion indices. For 
more details see the text.
\label{fig:zero}}
\end{figure}
One could think of similar Feynman graphs to diagrams~2 and 3 of fig.~\ref{fig:all_imcqc} but involving two-pions 
in one vertex from the Weinberg-Tomozawa term of ${\cal L}_{\pi N}^{(1)}$ eq.~\eqref{lags}.
These are depicted in fig.~\ref{fig:zero}. However, these diagrams are zero because of the antisymmetric isospin structure
of the Weinberg-Tomozawa vertex, proportional to $\varepsilon_{lm c}\tau^c$. The same type of pion is involved in the 
tadpole loop. This is clear for the diagram a) but also for diagram b) due to the diagonal structure of 
the vertex coupling the scalar source with two pions, eq.~\eqref{ver.l2}. As a result, both diagrams 
are zero when the pion indices are contracted with the antisymmetric tensor. 

The in-medium nucleon propagator reads \cite{fetter}, 
\be
G_0(k)_{i_3}=\frac{\theta(|\vk|-\xi_{i_3})}{k^0-E(\vk)+i\epsilon}+\frac{\theta(\xi_{i_3}-|\vk|)}{k^0-E(\vk)-i\epsilon}
            =\frac{1}{k^0-E(\vk)+i\epsilon}+2\pi i \, \delta(k^0-E(\vk)) \theta(\xi_{i_3}-|\vk|) ~,
\label{eq:nuc.pro.def}
\ee
with $E(\vk)=\vk^2/2 m$ the non-relativistic nucleon kinetic energy. 
In this equation the subscript $i_3$ is the third component of the
isospin of the nucleon, so that $i_3=+1/2$ corresponds to the proton and 
$i_3=-1/2$ to the neutron. The symbol $\xi_{i_3}$ is the Fermi momentum of the 
Fermi sea for the pertinent nucleon. 
We consider that in vacuum isospin symmetry is conserved, so that all the (vacuum) nucleon and pion masses are 
equal. The proton and neutron propagators can be combined in a common expression
\begin{equation}
 G_0(k) = \sum_{i_3} \Bigl( \frac{1}{2}+i_3\tau_3 \Bigr) G_0(k)_{i_3} ~.
\end{equation}

We make use of the Heavy Baryon $\chi$PT (HB$\chi$PT) pion-nucleon Lagrangian of \cite{fettes} up to ${\cal O}(p^2)$
\begin{align}
{\cal L}_{\pi N}^{(1)}&=\bar{N}\biggl[i D_0-\frac{g_A}{2}\vec{\sigma}\cdot \vec{u}\biggl]N~,\nn\\
{\cal L}_{\pi N}^{(2)}&=\bar{N}\biggl[
\frac{1}{2m}\vec{D}\cdot \vec{D} + c_1 \langle \chi_+ \rangle + \left(c_2-\frac{g_A^2}{8m}\right)u_0^2 + c_3 u_\mu u^\mu + c_5 \left( \chi_+ - \frac{\langle\chi_+\rangle}{2} \right) + \ldots \biggr]N
\label{lags}
\end{align}
where  $\langle \ldots \rangle$ denotes the trace in isospin space and the ellipses represent terms that are not needed here.
The covariant derivative is given by $D_\mu=\partial_\mu+\Gamma_\mu$, where the chiral connection is
$\Gamma_\mu=\frac{1}{2}[u^\dagger,\partial_\mu u]-\frac{i}{2}u^\dagger(v_\mu+a_\mu)u-\frac{i}{2}u(v_\mu-a_\mu)u^\dagger$.
The chiral vielbein is given by
$u_\mu=i\{u^\dagger,\partial_\mu u\}+u^\dagger(v_\mu+a_\mu)u-u(v_\mu-a_\mu)u^\dagger$.
The coupling of a scalar source to pions is also required. 
At lowest order in the chiral counting this coupling arises from
\begin{align}
{\cal L}_{\pi\pi}^{(2)}&= \frac{f_\pi^2}{4} \langle \chi_+ \rangle +\ldots
\label{pilags}
\end{align}
The matrix field of scalar sources $s(x)\equiv ||s_{ij}(x)||$,  is included in the operator $\chi_+$ \cite{gl1},
\begin{align}
\chi_+&= u^\dagger \chi u^\dagger + u \chi^\dagger u \, \notag \\
\chi&=2 B s(x) ~.
\end{align}
The parameter $B$ is related to the strength of the vacuum chiral quark condensate and the weak pion decay constant $f_\pi$, both in the chiral limit,  via
\be
 B\delta_{ij}=-\la 0|\bar{q}_iq_j|0\ra/f_\pi^2 ~.
 \label{vac.qc}
\ee
   The vertex coupling the scalar source $s_{ij}$  to two pions of Cartesian coordinates $a$ and $b$ can be evaluated
 straightforwardly from $-{\cal L}_{\pi\pi}^{(2)}$, eq.~\eqref{pilags}, with the result
\begin{align}
\label{ver.l2}
B\delta_{lm}\delta^{ab}~.
\end{align}
This vertex is diagonal both in the scalar source as well as in the pion fields.

\begin{figure}[ht]
\psfrag{k}{$k$}
\psfrag{p}{$p$}
\psfrag{l}{$\ell$}
\psfrag{pi}{$\pi$}
\psfrag{r}{$k-\ell$}
\centerline{\epsfig{file=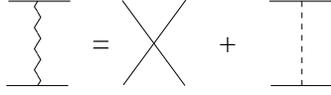,width=.25\textwidth,angle=0}}
\vspace{0.2cm}
\caption[pilf]{\protect \small
Leading order interaction kernel: the exchange of a wiggly line between two nucleons indicates in the following 
the sum of the ${\cal O}(p^0)$ local and the one-pion exchange contributions.
\label{fig:wig}}
\end{figure}

The nucleon-nucleon scattering amplitude at lowest order, ${\cal O}(p^0)$, is needed to match with our aim of calculating 
the NLO contributions to the chiral quark condensate in the nuclear medium. Note that the nucleon-nucleon scattering amplitudes 
 first appear in diagrams with $V_\rho=2$ so that 
it starts to give contributions to the quark condensate already at NLO.
These amplitudes originate from the Lagrangian 
with four nucleons, without quark masses or derivatives \cite{wein2}
\be
{\cal L}_{NN}^{(0)} = - \frac{1}{2}C_S(\bar{N}N)(\bar{N}N) - \frac{1}{2}C_T(\bar{N}\vec\sigma N)(\bar{N}\vec\sigma N) ~,
\label{lnn}
\ee
and from the one-pion exchange with the lowest order pion-nucleon coupling.
Its sum is represented diagrammatically in the following by the
exchange of a wiggly line as in fig.~\ref{fig:wig}. For explicit expressions of these amplitudes see \cite{Lacour:2009ej}, where 
we also compare with vacuum nucleon-nucleon scattering data. 

\section{$V_\rho=1$ contributions}
\label{sec:im_cqc.pc}

First we consider the contributions to the in-medium chiral quark condensate from pion-nucleon chiral dynamics.
 They are depicted in diagrams~1--3 of fig.~\ref{fig:all_imcqc}. 
For the evaluation of the different diagrams  
we need the vertex with the scalar source $s_{ij}$ coupling to a pair of nucleons $l$, $m$.
 It can be readily worked out from $-{\cal L}_{\pi N}^{(2)}$ 
eq.~\eqref{lags}, with the result
\begin{align}
-2B \bigl[ 2c_1\delta_{ij}\delta_{lm} + c_5\vec{\tau}_{ji}\cdot\vec{\tau}_{lm} \bigr] ~,
\label{eq:ver.sc.N}
\end{align}
where we have taken into account that $2\delta_{il}\delta_{j m}-\delta_{ij}\delta_{lm}=\vec{\tau}_{ji}\cdot\vec{\tau}_{lm}$. 
The diagram~1  then yields
\begin{equation}
 \Xi_1 = -2 B\left[2c_1 \delta_{ij}(\rho_p+\rho_n)+ c_5(\tau_3)_{ij}(\rho_p-\rho_n)\right] \doteq \Xi_1^{is} + \Xi_1^{iv} ~,
 \label{xi1}
\end{equation}
where the $\rho_p$ and $\rho_n$ are the proton and neutron densities given by $\rho_{p(n)}=\xi_{p(n)}^3/3 \pi^2$.
Notice that the isospin breaking contribution proportional to $c_5$ only involves the Pauli matrix $\tau_3$,
 so that for $i\neq j$ the contribution vanishes, as required. The contribution in eq.~\eqref{xi1} proportional to $c_1$ 
 is denoted in the following by  $ \Xi_1^{is}$ and that proportional to $c_5$  by $\Xi_1^{iv}$. 
The superscripts $is$ and $iv$ refer to the 
isoscalar and isovector character of these contributions, respectively.

\begin{figure}[ht]
\centerline{\epsfig{file=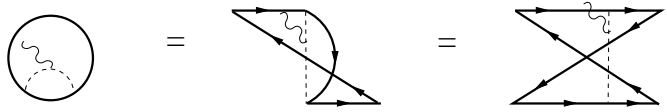,width=.5\textwidth,angle=0}}
\vspace{0.2cm}
\caption[pilf]{\protect \small The equivalence between diagram~3 and the 
crossed part of the one-pion exchange reduction of diagram~6 of fig.~\ref{fig:all_imcqc} is shown.
The diagram in the middle is an intermediate step in the continuous 
transformation of the diagram on the left-hand-side to the one on the 
right-hand-side.
\label{fig:equiv_cqc}}
\end{figure}

Let us proceed to the evaluation of the NLO diagrams 2 and 3 of fig.~\ref{fig:all_imcqc}.  
Diagram~2  originates by dressing the in-medium nucleon propagator with the one-pion loop
nucleon self-energy, $ \Sigma^\pi$. It is given by
\begin{equation}
\Xi_2 =- 2iB \int\frac{d^4k}{(2\pi)^4} e^{ik^0\eta} \hbox{ Tr} \left\{ \bigl[ 2 c_1 \delta_{ij} + c_5 \vec\tau_{ji}\cdot\vec\tau \bigr] G_0(k) \, \Sigma^\pi \, G_0(k) \right\}  \doteq \Xi_2^{is} + \Xi_2^{iv}~,
\label{xi2.1}
\end{equation}
with the convergence factor $e^{ik^0\eta}$, $\eta\rightarrow0^+$, associated with any closed loop made up by a single nucleon line \cite{fetter}.
The trace, indicated by Tr, acts both in spin and isospin spaces. Similarly as in eq.~\eqref{xi1} the term proportional to $c_1$ is denoted by 
$\Xi_2^{is}$ and that  proportional to $c_5$ by $\Xi_2^{iv}$. 
The contribution from diagram~2 with only free parts in all the  nucleon propagators involved, including that in $\Sigma^\pi$,
  vanishes.  This is discussed in detail in ref.~\cite{Lacour:2009ej} for similar diagrams that appear in the calculation of 
the  in-medium pion self-energy and nuclear matter energy. Briefly, it follows just by  closing the integration contour  
on the complex $k^0$ half-plane opposite to that where the poles lie.
 Another important point to keep in mind is that the contributions with only Fermi sea insertions 
in all nucleon propagators is part of the $V_\rho=2$ contribution of the crossed exchange part of diagram~4 in fig.~\ref{fig:all_imcqc}.
This is shown diagrammatically in 
fig.~3 of \cite{Lacour:2009ej},
where the two external pion sources should be replaced by the scalar source for the case at hand (similarly, see fig.~\ref{fig:equiv_cqc} for diagram 4.)  
The different $V_\rho=2$ contributions are evaluated in section \ref{sec:im_cqc.vrho2}.
Consequently, only those terms that involve simultaneously  free-space as well as density-dependent
 parts of the nucleon propagators 
are considered here. For the isovector part the calculation of the contribution $\Pi_5$ in ref.~\cite{Lacour:2009ej} 
applies straightforwardly, just by replacing the vertex function. The same procedure applied to the isoscalar part 
drives to its cancellation, $\Xi_2^{is}=0$. For the same reason as in \cite{Lacour:2009ej} $\Xi_2^{iv}$ turns to be 
of ${\cal O}(p^6)$ or N$^2$LO.  This is due to the appearance of the derivative of the free part of the nucleon one-pion loop, 
$\Sigma_f^\pi$, with respect to energy. This derivative is finally suppressed by one chiral order \cite{Lacour:2009ej}.

For diagram~3 in fig.~\ref{fig:all_imcqc}  the scalar source is attached to the pion propagator involved in the one-pion loop nucleon self-energy. 
In the same way as discussed above for diagram~2,  the contribution with only the free part for all nucleon propagators 
vanishes, while that with all nucleon propagators as in-medium insertions are accounted 
for by the exchange part of diagram~6 of fig.~\ref{fig:all_imcqc}, see fig.~\ref{fig:equiv_cqc}. This is calculated in section~\ref{sec:im_cqc.vrho2}. 
Then, we consider here the part of diagram 3 having one free nucleon propagator and the density dependent part 
of the other. Let us recall that the coupling of a scalar source to two pions was already given in eq.~\eqref{ver.l2}.  
There is a close relationship between diagram~3 of fig.~\ref{fig:all_imcqc} and the contribution to the nuclear matter energy 
from the one-pion loop nucleon self-energy, diagram 2 in  fig.~17 of \cite{Lacour:2009ej},  denoted by ${\cal E}_2$. Of course, this is a requirement from the Hellmann-Feynman theorem. 
It is straightforward to check that
\begin{align}
\Xi_3 &= iB\delta_{ij} \frac{1}{2} \int\frac{d^4k}{(2\pi)^4} e^{ik^0\eta} \hbox{Tr}\left[
 G_0(k) \frac{\partial \Sigma^\pi(k)}{\partial m_\pi^2}\right]
 = B\delta_{ij}\frac{\partial {\cal E}_2}{\partial m_\pi^2}~,
\label{dmpi2.sig3}
\end{align}
with the derivative affecting only the explicit dependence of $\Sigma_f^\pi$ on the pion propagator,
 and not including the implicit one from the nucleon mass dependence on it. One has to subtract the 
value of the one-pion loop nucleon self-energy at $k^0=0$ since we are using the physical nucleon mass.
After performing the $k^0$ integration, one has the expression
\begin{align}
\Xi_3&=-2B\delta_{ij}\int\frac{d^3k}{(2\pi)^3}\Bigl( \theta(\xi_p-|\vk|)+\theta(\xi_n-|\vk|) \Bigr)
 \frac{\partial \Sigma^\pi_f(\omega)}{\partial m_\pi^2}~,
\label{xi3}
\end{align}
where $ \omega=E(\vk)$ and the partial derivative is given by
\begin{align}
\frac{\partial \Sigma^\pi_f(\omega)}{\partial m_\pi^2}&=
\frac{3 g_A^2}{64 \pi^2 f_\pi^2 m_\pi^2}\left[
2 \omega^3 - 4\omega m_\pi^2 - 3\pi m_\pi^3 + 3 m_\pi^2\sqrt{b}\left(\pi+
i\ln\frac{\omega+i\sqrt{b}}{-\omega+i\sqrt{b}}\right)
\right]~,
\end{align}
with $b=m_\pi^2-\omega^2-i\epsilon$ and $\epsilon\to 0^+$. Notice that $\Xi_3$ is an ${\cal O}(p^7)$ or N$^2$LO
 contribution because $\partial \Sigma_f^\pi/\partial m_\pi^2={\cal O}(p^2)$. The ${\cal O}(p)$ contribution from the 
term $-3 \pi m_\pi$ is cancelled by that coming from $3 \pi \sqrt{b}$. 
We then conclude that the only contribution with $V_\rho=1$ 
up-to-and-including NLO is given by $\Xi_1$, eq.~\eqref{xi1}.

\section{$V_\rho=2$ contributions}
\label{sec:im_cqc.vrho2}

We now consider those next-to-leading order contributions to the in-medium chiral quark condensate
 that involve the nucleon-nucleon interactions. They are depicted in diagrams~4--6 of the last two rows
 of fig.~\ref{fig:all_imcqc}. 

\subsection{Contributions $\Xi_4$ and $\Xi_5$}
\label{sub:im_cqc.xi4.5}

Diagrams~4 and 5  are analogous to  diagrams~a) and c) of fig.4 in \cite{nlou}.
There it has been shown that those diagrams cancel each other.
This argument was developed in
\cite{Lacour:2009ej}
within a specific method for resumming the iteration of the wiggly lines for the non-perturbative nucleon-nucleon interactions.
We expect that this cancellation also takes place for the case of the in-medium chiral quark condensate.
However, the vertex coupling two nucleons with the scalar source, eq.~\eqref{eq:ver.sc.N}, has both an isoscalar
 and an isovector term while for the aforementioned cancellation in the case of the pion self-energy only an isovector vertex was involved.
Here we want to show on general grounds that the cancellation also takes place for the in-medium chiral quark condensate,
 following similar arguments as those in \cite{nlou}. 

\begin{figure}[ht]
\psfrag{k1}{$k_1$}
\psfrag{k2}{$k_2$}
\psfrag{k1-k}{$k_1-q$}
\psfrag{k2+k}{$k_2+q$}
\centerline{\epsfig{file=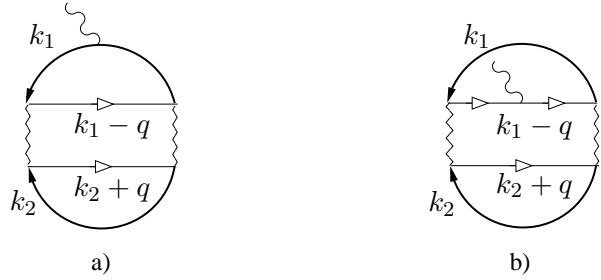,width=.45\textwidth,angle=0}}
\vspace{0.2cm}
\caption[pilf]{\protect \small
Contribution to the chiral quark condensate with a two-nucleon reducible loop. 
The scalar source couples outside the loop for diagram~a) and inside it for diagram~b).
\label{fig:cancel1loop}}
\end{figure} 

Diagram~4 of fig.~\ref{fig:all_imcqc} can be written in terms of the nucleon self-energy due to 
the in-medium nucleon-nucleon scattering, denoted by $\Sigma^{NN}$. It reads
\begin{align}
\Xi_4 &= -2iB \int\frac{d^4k_1}{(2\pi)^4} e^{ik^0_1\eta} \hbox{ Tr} 
\Bigl\{ \bigl(2 c_1\delta_{ij}+ c_5 \vec{\tau}_{ji}\cdot\vec{\tau}\bigr) G_0(k_1)\Sigma^{NN}G_0(k_1) \Bigr\} ~.
\label{xi4.1}
\end{align}
The expression for $\Sigma_{\alpha_1,NN}$, corresponding to the self-energy of a nucleon with isospin $\alpha_1$, is  \cite{nlou} 
\begin{align}
\Sigma_{\alpha_1,NN}=-i\sum_{\alpha_2,\sigma_2}\int\frac{d^4 k_2}{(2\pi)^4}G_0(k_2)_{\alpha_2} 
T_{\alpha_1\alpha_2}^{\sigma_1\sigma_2}(k_1,k_2) e^{ik_2^0\eta}~.
 \label{sel.n}
\end{align}
Here $\alpha_2,$ $\sigma_2$ correspond to the running third components of isospin and spin, respectively, while $\sigma_1$ is 
the third component  of the spin of the external nucleon.  In addition, $T_{\alpha_1\alpha_2}^{\sigma_1\sigma_2}(k_1,k_2)$
 refers to the elastic  in-medium  nucleon-nucleon scattering 
amplitude for $N_{\alpha_1,\sigma_1}(k_1)N_{\alpha_2,\sigma_2}(k_2)\to N_{\alpha_1,\sigma_1}(k_1) N_{\alpha_2,\sigma_2}(k_2)$. 
We also make use of the identity $\partial G_0(k)/\partial k^0=-G_0(k)^2.$ As a result, eq.~\eqref{xi4.1} can be expressed as 
\begin{align}
\Xi_4 &= -2B \sum_{\alpha_1,\alpha_2} \sum_{\sigma_1,\sigma_2} \int\frac{d^4k_1}{(2\pi)^4}\frac{d^4k_2}{(2\pi)^4} e^{ik^0_1\eta} e^{ik^0_2\eta} 
 G_0(k_1)_{\alpha_1}  \bigl( 2c_1\delta_{ij}+c_5(\tau_3)_{ij} (\tau_3)_{\alpha_1\alpha_1} \bigr) 
G_0(k_2)_{\alpha_2}  \frac{\partial T^{\sigma_1\sigma_2}_{\alpha_1\alpha_2}(k_1,k_2)}{\partial k_1^0}~,
 \label{xi4.2}
\end{align}
where in the last equation an integration by parts in $k_1^0$ has been performed. 
In order to see the cancellation between diagrams~4 and 5 of fig.~\ref{fig:all_imcqc} let us proceed 
similarly as in ref.~\cite{nlou}, for the 
case of the in-medium pion self-energy, and consider first the case with only one
 reducible two-nucleon diagram, fig.~\ref{fig:cancel1loop}.
The contribution of  diagram a), $\Xi_4^L$, can be readily worked from  
eq.~\eqref{xi4.2} by reducing $T^{\sigma_1\sigma_2}_{\alpha_1\alpha_2}(k_1,k_2)$ to its one-loop calculation. It results
\begin{align}
\Xi^L_4&= -2B \sum_{\alpha_1,\alpha_2} \sum_{\sigma_1,\sigma_2} \int\frac{d^4k_1}{(2\pi)^4}\frac{d^4k_2}{(2\pi)^4} e^{ik^0_1\eta} e^{ik^0_2\eta}
G_0(k_1)_{\alpha_1} \bigl( 2 c_1 \delta_{ij} + c_5(\tau_3)_{ij} (\tau_3)_{\alpha_1\alpha_1} \bigr) G_0(k_2)_{\alpha_2} \nn\\
& \times \frac{\partial}{\partial k_1^0} \biggl[ \frac{-i}{2} \sum_{\alpha'_1,\alpha'_2} 
 \int\frac{d^4q}{(2\pi)^4} V_{\alpha_1 \alpha_2;\alpha'_1\alpha'_2}(q) G_0(k_1-q)_{\alpha'_1}  G_0(k_2+q)_{\alpha'_2}
 V_{\alpha'_1 \alpha'_2;\alpha_1\alpha_2}(-q) \biggr]~,
\label{xi4.3}
\end{align}
where $V_{\alpha\beta;\gamma\delta}$ corresponds to the wiggly line with the first pair of labels belonging to the outgoing nucleons
 and the second pair to the in-going ones.
Note that a symmetry factor $1/2$ is included because $V$ contains both the direct and exchange terms. In addition, $V$ 
also will depend generally on spin. 
Both isospin and spin indices are globally indicated in the labels of $V$ with Greek letters. The quantity in squared brackets in the 
previous equation is $T_{\alpha_1\alpha_2}^{\sigma_1\sigma_2}(k_1,k_2)$ at the one-loop level. It is not necessary to consider
 $T_{\alpha_1\alpha_2}^{\sigma_1\sigma_2}$ at the tree-level, consisting of the exchange of one wiggly line, fig.~\ref{fig:wig},
 because it is then independent on $k_1^0$ so that the derivative with respect to $k_1^0$ is zero.    

 One can similarly write down the contribution from the diagram b) of  fig.~\ref{fig:cancel1loop}, denoted by $\Xi_5^L$. It reads
\begin{align}
\Xi^L_5&= 2B \sum_{\alpha_1,\alpha_2} \sum_{\sigma_1,\sigma_2} \int\frac{d^4k_1}{(2\pi)^4}\frac{d^4k_2}{(2\pi)^4} e^{ik^0_1\eta} e^{ik^0_2\eta}
G_0(k_1)_{\alpha_1}  G_0(k_2)_{\alpha_2} \frac{\partial}{\partial k_1^0} 
\biggl[ \frac{-i}{2} \sum_{\alpha'_1,\alpha'_2} 
 \int\frac{d^4q}{(2\pi)^4} V_{\alpha_1 \alpha_2;\alpha'_1\alpha'_2}(q) \nn\\
&\times \bigl( 2 c_1 \delta_{ij} + c_5(\tau_3)_{ij} (\tau_3)_{\alpha'_1\alpha'_1} \bigr)
 G_0(k_1-q)_{\alpha'_1}  G_0(k_2+q)_{\alpha'_2}
 V_{\alpha'_1 \alpha'_2;\alpha_1\alpha_2}(-q) \biggr]~,
\label{xi5}
\end{align}
with the global symmetry factor $1/2$ from closing the lines.
The appearance of the derivatives with respect to $k_1^0$ is again due to the fact that the propagator attached to the scalar source appears squared.
The loop integrals between the squared brackets in eqs.~\eqref{xi4.3} and \eqref{xi5} are typically divergent.
Nevertheless, the parametric derivative with respect to $k_1^0$ can be extracted out of the integral of eq.~\eqref{xi5} as soon as it is regularized.
Summing over all isospin states makes clear that the position of the vertex associated with the coupling of the scalar source to 
two nucleons, either inside or outside the squared brackets, does not yield a difference. For that one has to keep in mind that 
$T_{\alpha_1\alpha_2}^{\sigma_1\sigma_2}(k_1,k_2)=T_{\alpha_2\alpha_1}^{\sigma_2\sigma_1}(k_2,k_1)$, due to the  Fermi-Dirac statistics for a pair 
of two-nucleon states. Furthermore, notice that all the indices and four-momenta associated with the nucleons 1 and 2 
are summed and integrated, respectively.
In this way eqs.~\eqref{xi4.3} and \eqref{xi5} mutually cancel.\footnote{Alternatively, note that the term proportional to $c_1$ in eq.~\eqref{xi5} is the same independently of its location because is proportional 
to the identity matrix. That part proportional to $c_5$ is purely isovector, as those terms stemming from the Weinberg-Tomozawa vertex in refs.~\cite{nlou,Lacour:2009ej}. As shown there only the states with $I_3=\pm 1$ give contribution so that $c_5(\tau_3)_{\alpha'_1\alpha'_1}$ in eq.~\eqref{xi5} gives rise to $c_5 I_3$, which then trivially factorizes for the separate contributions of the proton-proton and neutron-neutron states.}

\begin{figure}[ht]
\centerline{\epsfig{file=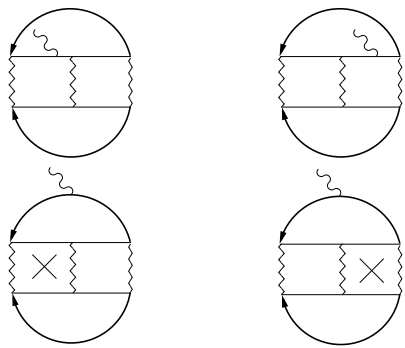,width=.3\textwidth,angle=0}}
\vspace{0.2cm}
\caption[pilf]{\protect \small
After performing the integration by parts in eq.~\eqref{xi4.2} the derivative with respect to $k_1^0$ acts onto the scattering amplitude.
This gives a sum of derivatives acting on two-nucleon reducible loops, indicated by the crosses.
When the derivative acts on a baryon propagator the latter becomes squared.
In this way, the first diagram in the second row of the figure equals the one of the first row but with opposite sign and they cancel each other.
The same applies to the second diagrams in both rows.
\label{fig:cancel2loop}}
\end{figure}

This process of mutual cancellation can be generalized to any number of two-nucleon reducible loops, using the same argument as given in \cite{nlou} 
for the case of the in-medium pion self-energy. 
An $n+1$ iterated wiggly line exchange implies $n$ two-nucleon reducible loops.
The scalar source can be attached to any of them for $\Xi_5$, while for $\Xi_4$ the derivative with respect to $k_1^0$ can also act 
on any of the loops. This is exemplified in fig.~\ref{fig:cancel2loop} for the case with two two-nucleon reducible loops. Hence,
\begin{equation}
\Xi_4+\Xi_5=0 ~.
\label{cancel.45}
\end{equation}
The basic simple reason for such cancellation is that while for $\Xi_5$ there is a derivative acting onto the scattering amplitude,
 $\Xi_4$ involves an integration by parts in order to do so, which then introduces an extra minus sign.

The previous cancellation is also explicitly obtained making use of Unitary $\chi$PT applied to nuclear matter as 
developed in ref.~\cite{Lacour:2009ej}.
The intermediate result of \cite{Lacour:2009ej} can be used straightforwardly by considering that instead of the 
Weinberg-Tomozawa vertex and pion-nucleon Born terms, used in the problem 
of the in-medium pion self-energy, one has the nucleon vertex of eq.~\eqref{eq:ver.sc.N} with the scalar source.
The partial wave decomposition of $\Xi_4$ reads
\begin{align}
 \Xi_4 &= -2B \sum_{J,\ell,S,I}\sum_{\alpha_1,\alpha_2}(2J+1) \chi(S \ell I)^2 \int\frac{d^4 k_1}{(2\pi)^4}\frac{d^4 k_2}{(2\pi)^4} e^{ik^0_1\eta} e^{ik^0_2\eta} 
          \bigl[ 2c_1\delta_{ij} + I_3c_5(\tau_3)_{ij} \bigr] \nn\\
       &\quad\times G_0(k_1)_{\alpha_1} G_0(k_2)_{\alpha_2} m\frac{\partial T_{JI}^{I_3}(\ell,\ell,S)}{\partial A}   ~,
\label{xi4.pw}
\end{align}
with $\alpha_1+\alpha_2=I_3$, the third component of the total isospin $I$ of the two-nucleon state. Other symbols 
used are  $J$ the total angular momentum, $\ell$ the orbital angular momentum and $S$ the total spin of the nucleon-nucleon pair. We also
 employ the kinematic variable $A=2m a^0-\mathbf{a}^2$, where $a=(k_1+k_2)/2$ and $\mathbf{a}$ is the three-vector made by 
the spatial components of $a$. For on-shell scattering $A=\vp^2$, with $\vp$ the three-momentum in the two-nucleon rest frame.    
For the term proportional  to $c_5$ due to its isovector character   only the difference between the proton-proton 
and neutron-neutron contributions survives ($I_3=0$ for proton-neutron pairs).
For the expressions at ${\cal O}(p^6)$ and ${\cal O}(p^7)$ we plug  into eq.~\eqref{xi4.pw} the derivative $\partial T_{JI}/\partial A$ at leading and next-to-leading order, according to 
 eqs.~(5.24) and (5.25) of \cite{Lacour:2009ej}, respectively.
For the contribution $\Xi_5$ the general partial wave decomposition yields
\begin{align}
\Xi_5 = -\frac{1}{2} \sum_{J,\ell,S,I}\sum_{\alpha_1,\alpha_2}(2J+1)
\chi(S \ell I)^2 \int\frac{d^4k_1}{(2\pi)^4}\frac{d^4k_2}{(2\pi)^4} e^{ik^0_1\eta} e^{ik^0_2\eta}
 G_0(k_1)_{\alpha_1} G_0(k_2)_{\alpha_2} \left[D_{JI}^{I_3}\right]^{-1}\cdot\xi_{JI}^{I_3}\cdot \left[D_{JI}^{I_3}\right]^{-1}~.
\label{xi5.pw}
\end{align}
The expressions for $D_{JI}^{I_3}$ at LO and NLO are  given in
 eq.~(5.25) of \cite{Lacour:2009ej}
and $\xi_{JI}^{I_3}$ is calculated order by order.
For that one has to work out the partial wave decomposition of the two-nucleon reducible diagram with the scalar source attached to one of the nucleon propagators inside the loop.
Following those calculations one obtains for the leading order (LO) and next-to-leading order (NLO)
\begin{align}
\xi_{JI}\bigr|_{LO~~} &= -\bigl[N_{JI}^{(0)}\bigr]^2 \cdot DL_{10}~,\nn\\
\xi_{JI}\bigr|_{NLO} &= DL_{JI}^{(1)}-\left\{L_{JI}^{(1)}+\bigr[N_{JI}^{(0)}\bigl]^2 L_{10},N_{JI}^{(0)}\right\}DL_{10}~,\nn\\
DL_{10}                     &= -4 B \bigl[ 2c_1\delta_{ij} + I_3c_5(\tau_3)_{ij} \bigr] \frac{m\partial L_{10}}{\partial A}  ~, \nn\\
DL_{JI}^{(1)}               &= -4 B \bigl[ 2c_1\delta_{ij} + I_3c_5(\tau_3)_{ij} \bigr] \frac{m\partial L_{JI}^{(1)}}{\partial A}  ~.
\label{xi5.lo}
\end{align}
The contribution $\Xi_5$ at ${\cal O}(p^6)$ and ${\cal O}(p^7)$ results when the just given expressions for $\xi_{JI}$, in order, are inserted in 
eq.~\eqref{xi5.pw}. It is straightforward to check that they exactly cancel with $\Xi_4$ accordingly.

\subsection{Contribution $\Xi_6$}
\label{sub:im_cqc.xi6}

Let us now consider the calculation of diagram~6 of fig.~\ref{fig:all_imcqc}, where the scalar source is attached to an exchanged wiggly line.
Since the scalar source coupling to a local term is of higher order, it only couples to the pion exchange lines here. 
The appearance of the pion propagator squared with a zero momentum scalar source leads to its derivative with respect to the pion mass.
The local term is independent of the pion mass, so we may act the derivative onto the explicit dependence    on the pion-mass of the
 full nucleon-nucleon scattering amplitude. The implicit dependence  of the latter on $m_\pi^2$ due to  
that of the nucleon mass is the content of the diagrams 4 and 5 of fig.~\ref{fig:all_imcqc}, which have been shown   above to cancel mutually. We can write 
\begin{align}
\Xi_6&
= \frac{1}{2}B \delta_{ij}  \sum_{\alpha_1,\alpha_2}\sum_{\sigma_1,\sigma_2}
\int\frac{d^4k_1}{(2\pi)^4}\frac{d^4k_2}{(2\pi)^4} e^{ik^0_1\eta} e^{ik^0_2\eta} G_0(k_1)_{\alpha_1}
 G_0(k_2)_{\alpha_2} \frac{\partial  T^{\sigma_1\sigma_2}_{\alpha_1\alpha_2}(k_1,k_2)}{\partial m_\pi^2}
=B \delta_{ij}\frac{\partial {\cal E}_3}{\partial m_\pi^2} ~.
\label{xi6}
\end{align}
Where ${\cal E}_3$ is the contribution to the nuclear matter energy due to the
in-medium nucleon-nucleon interaction at leading order, evaluated in 
ref.~\cite{Lacour:2009ej}. In order to proceed we have to evaluate the derivative of the nucleon-nucleon scattering amplitude with respect to $m_\pi^2$.
We decompose the in-medium nucleon-nucleon scattering amplitude in a sum over nucleon-nucleon partial waves $T_{JI}(\ell',\ell,S)$, 
as previously done for $\Xi_4$, eq.~\eqref{xi4.pw}. We distinguish now between $\ell$ and $\ell'$, corresponding to the 
 the initial and final orbital angular momentum, in this order. 
In terms of the nucleon-nucleon interaction kernel $N_{JI}(\ell',\ell,S)$, the nucleon-nucleon partial wave $T_{JI}$ in Unitary $\chi$PT is given by \cite{Lacour:2009ej}
\begin{equation}
 T_{JI}(\ell',\ell,S) = \left[ 1 + N_{JI}(\ell',\ell,S) \cdot L_{10} \right]^{-1} \cdot N_{JI}(\ell',\ell,S)
\equiv D_{JI}^{-1}\cdot N_{JI} ~,
\label{tji.1}
\end{equation}
with $L_{10}$ the nucleon-nucleon unitarity scalar function \cite{Lacour:2009ej}. This function does not depend explicitly 
on the pion mass.   We have also introduced  $D_{JI}= 1 + N_{JI} \cdot L_{10}$. 
Eq.~\eqref{tji.1} can also be rewritten as
\begin{equation}
 T_{JI} = N_{JI} - N_{JI} \cdot L_{10} \cdot T_{JI} ~.
\end{equation}
Taking the derivative with respect to the explicit dependence on $m_\pi^2$ on
both sides of the previous expression leads to
\begin{align}
\frac{\partial T_{JI}}{\partial m_\pi^2}&= \left[D_{JI}\right]^{-1} \cdot  \frac{\partial N_{JI}}{\partial m_\pi^2} 
 \cdot \left[D_{JI}\right]^{-1} ~.
\label{derivative.t}
\end{align}

In terms of this equation the partial wave decomposition of $\Xi_6$ at NLO reads 
\begin{align}
\Xi_6 &= \frac{1}{2} B\delta_{ij} \sum_{J,\ell,S,I}\sum_{\alpha_1,\alpha_2} (2J+1) \chi(S \ell I)^2
\int\frac{d^4 k_1}{(2\pi)^4}\frac{d^4 k_2}{(2\pi)^4} e^{ik^0_1\eta} e^{ik^0_2\eta} 
G_0(k_1)_{\alpha_1} G_0(k_2)_{\alpha_2} \frac{\partial T_{JI}^{I_3}}{\partial m_\pi^2}\biggr|_{LO}  ~.
\label{xi.c.2}
\end{align}
On the problem of evaluating and regularizing this amplitude we refer to the related detailed discussion in \cite{Lacour:2009ej} for the calculation of ${\cal E}_3$.

\section{Discussion and results}
\label{sec:im_cqc.num}

It follows from our calculation for the in-medium corrections shown in fig.~\ref{fig:all_imcqc} that  the chiral quark condensate up-to-and-including NLO is given by
\begin{equation}
m_q  \langle \Omega|\bar{q}_iq_j|\Omega\rangle = m_q \langle 0|\bar{q}_iq_j|0 \rangle + m_q (\Xi_1 + \Xi_6) ~.
\label{xi.nlo}
\end{equation}
Here we have taken into account that $\Xi_2$ and $\Xi_3$ are indeed one order higher (NNLO), while $\Xi_4+\Xi_5=0$. 
 Both $\Xi_1$ and $\Xi_6$ are clearly connected with the corresponding contribution to the 
nuclear matter energy, as required by the Hellmann-Feynman theorem \cite{Drukarev:1991fs,hell1,lutz},
\begin{equation}
 m_q \la\Omega|\bar{q}_iq_j|\Omega\ra - m_q \la0|\bar{q}_i q_j|0\ra 
= \frac{m_q}{2} \left( \delta_{ij} \frac{d }{d \hat{m}}+(\tau_3)_{ij}\frac{d}{d\bar{m}}\right)(\rho\, m+{\cal E}) 
 ~,
\label{mayonesa}
\end{equation}
with $\bar{m}=(m_u-m_d)/2$ and ${\cal E}$ the energy density of the nuclear matter system. 
This constraint is fulfilled in our case, where $\Xi_1$,  eq.~\eqref{xi1},  is the leading 
derivative with respect to $ m_\pi^2$ of the nucleon mass.\footnote{At lowest
  order in the chiral expansion $m_\pi^2=2 B \hat{m}$ so that in eq.~\eqref{mayonesa} 
one can make use of the operator $m_\pi^2 d/dm_\pi^2$ instead of $\hat{m}
d/d\hat{m}$.} 
In turn, $\Xi_6$  corresponds to the explicit derivative of 
the  nuclear matter energy density due to the nucleon-nucleon interactions
with respect to $m_\pi^2$, eq.~\eqref{xi6}. Notice, that the 
implicit dependence on $m_\pi^2$ of the in-medium nucleon-nucleon interactions
does not give 
any contribution to $\Xi$ because of 
the discussed above mutual cancellation between $\Xi_4$ and $\Xi_5$ at NLO.

The term $\Xi_1^{is}$ can be written directly in terms of the pion-nucleon 
$\sigma$ term with the result
\begin{align}
\Xi_1^{is}&=-\la 0|\bar{q}_i q_j|0\ra \frac{\sigma (\rho_p+\rho_n) }{f_\pi^2 m_\pi^2}~,
\label{xi1.f}
\end{align}
using the Gell-Mann--Oakes--Renner relation (GMOR) \cite{renner}, 
$m_\pi^2 f_\pi^2=-2 \hat{m}\la 0|\bar{q}_i q_j|0\ra$, 
valid at lowest order in the chiral expansion, see e.g.~\cite{ulfrep0}.
Regarding the numerical value of $\sigma$ one has  the earlier extraction 
$\sigma=45\pm 8$~MeV of \cite{Gasser:1990ap,Koch:1982pu} or
 the one from the more recent partial wave analysis of pion-nucleon 
scattering \cite{Pavan:2001wz} $\sigma=64\pm 7$~MeV.
In both cases the dispersion analysis of the nucleon scalar form factor of 
\cite{Gasser:1990ap} to estimate the departure
 between the sigma-term and the $\pi N$ scattering amplitude evaluated at the 
(unphysical) Cheng-Dashen point is used. In asymmetric nuclear matter there are
additional isospin symmetry breaking terms proportional to $c_5$ stemming from diagrams~1 and 2, giving 
rise to the NLO contribution $\Xi_1^{iv}$, eq.~\eqref{xi1}.
This contribution, which distinguishes between the $\bar{u}u$ and $\bar{d}d$
quark condensates, are suppressed because they are proportional to the
numerically small low-energy constant $c_5=-0.09\pm 0.01~$GeV$^{-1}$ \cite{aspects}. 
For symmetric nuclear matter this contribution is zero since it is
proportional to the difference of proton and neutron densities.

\begin{figure}[ht]
\psfrag{rho}{\small $\begin{array}{c}\\\rho~(\hbox{fm}^{-3})\end{array}$}
\centerline{\epsfig{file=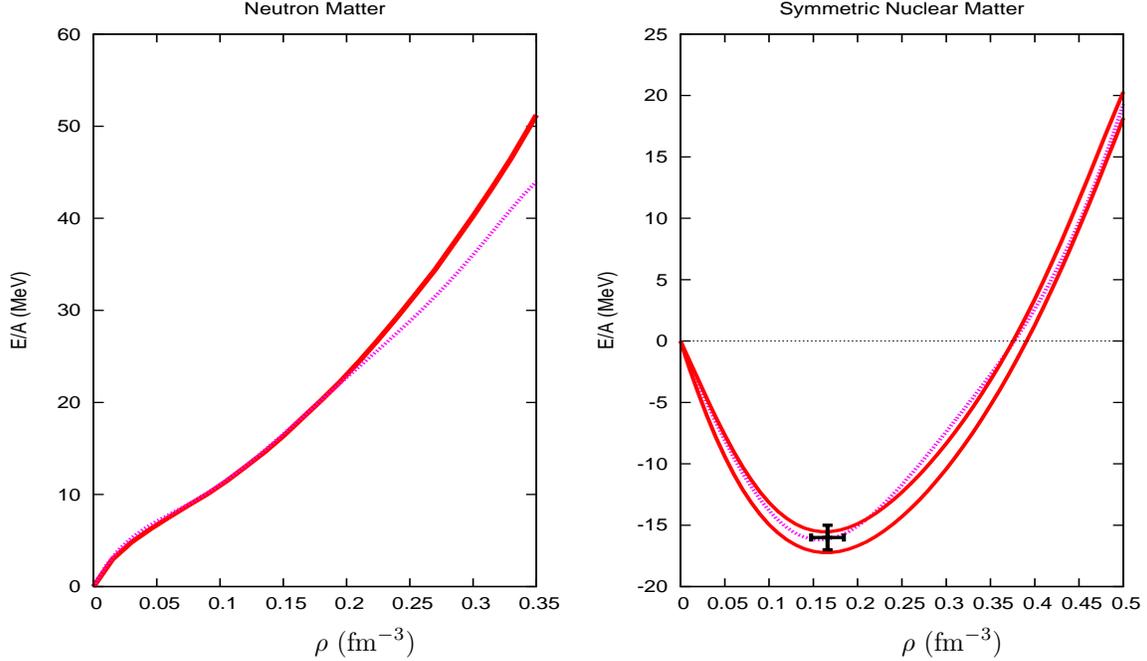,height=.7\textheight,width=.5\textwidth,angle=-90}}
\vspace{0.2cm}
\caption[pilf]{\protect \small (Color online.) ${\cal E}/\rho$ for neutron
  matter (left panel) and for symmetric  nuclear matter (right panel). 
The two (magenta) dotted lines correspond 
to ref.~\cite{urbana} and the solid ones are our calculation.
\label{fig:energy}}
\end{figure}
 
 The energy density and energy per particle, ${\cal E}/\rho$,  have been calculated in
\cite{Lacour:2009ej}. In this reference, saturation of symmetric nuclear
matter is obtained from the consistent chiral power counting, eq.~\eqref{fff}. 
In addition,
a remarkably good agreement with sophisticated many-body calculations
\cite{panda} was obtained for the equation of 
state for neutron matter
in terms of just one free parameter, a subtraction constant called $g_0$. The
appearance of this subtraction constant is due to the purely free part of 
the scalar two-nucleon reducible loop, $L_{10}$ already introduced in 
eq.~\eqref{tji.1}. Its free part, denoted by $g(A)$, is a divergent 
function which requires a subtraction \cite{Lacour:2009ej}. 
Taking $D$ as the subtraction point one has
\begin{eqnarray}
g(A)&=&g(D)-\frac{m(A-D)}{4\pi^2}\int_0^\infty
dk^2\frac{k}{(k^2-A-i\epsilon)(k^2-D-i\epsilon)}\nonumber\\
&=&g(D)-\frac{i m}{4\pi}\left( \sqrt{A}-i\sqrt{|D|}\right)=
g_0-i\frac{m\sqrt{A}}{4\pi}~.
\end{eqnarray}
${\cal E}_3$, eq.~(7.36) in \cite{Lacour:2009ej}, depends on $g_0$ both implicitly, due to the dependence   
of the partial wave amplitudes $T_{JI}(\ell',\ell,S)$ on  $g_0$,
cf.eq.~\eqref{tji.1},  and explicitly, in a linear manner.  For the latter 
dependence the symbol $\widetilde{g}_0$ was introduced in 
ref.~\cite{Lacour:2009ej} (to which we refer for a more detailed discussion on 
$g_0$ and $\widetilde{g}_0$). 
For pure neutron matter both $g_0$ and $\widetilde{g}_0$ are obtained with the 
value $g_0=\widetilde{g}_0\simeq -0.6~m_\pi^2$, very close to their expected 
natural size~$g_0\simeq -0.5$~$m_\pi^{-2}$. 
The case of symmetric nuclear matter requires some fine-tuning of $g_0$, with a 
final value of $g_0\simeq - m_\pi^2$, while $\widetilde{g}_0$ is kept on its 
expected value of $-0.5~m_\pi^2$. 
With that, results in perfect agreement with experiments 
were obtained for the saturation density, energy per particle and compression 
modulus \cite{Lacour:2009ej}, see fig.~\ref{fig:energy} where the equations 
of state for neutron (left panel) and symmetric nuclear matter (right panel) 
are shown.  The magenta lines are from the many-body calculations of refs.~\cite{urbana,panda} 
employing to so-called realistic nucleon-nucleon potentials. The agreement
between these references and our results is remarkable. 
We use our previous evaluation of ${\cal E}_3$ for
calculating $\Xi_6$, eq.~\eqref{xi6}, by taking the derivative with respect to
the  explicit dependence of ${\cal  E}_3$  on $m_\pi^2$.  It is important to
stress that the in-medium quark condensate does not depend on
$\widetilde{g}_0$ because this parameter multiplies a quantity which is
independent of the pion mass,  denoted by $\Sigma_{\infty \ell}$ in 
ref.~\cite{Lacour:2009ej}. The latter originates from the limit
$\vp^2\to\infty$ of the one-pion-exchange tree level partial waves at lowest
order and this limit is independent of $m_\pi^2$.   Making use again  of the
 GMOR relation, $\Xi_6$ eq.~\eqref{xi6}  can be written as
\begin{align}
\Xi_6=-\la 0|\bar{q}_iq_j|0\ra\frac{1}{f_\pi^2}\frac{\partial {\cal E}_3}{\partial m_\pi^2}(1+{\cal O}(m_\pi^2))~.
\label{xi6.f}
\end{align}
Putting together the previous equation with $\Xi_1^{is}$, eq.~\eqref{xi1.f}, 
and  $\Xi_1^{iv}$, eq.~\eqref{xi1}, the ratio between the in-medium and vacuum 
quark condensate to NLO reads
\begin{align}
\frac{\la \Omega|\bar{q}_iq_j|\Omega\ra}{\la 0|\bar{q}_i q_j|0\ra}&=1-\frac{\sigma (\rho_p+\rho_n) }{f_\pi^2 m_\pi^2}+\frac{2 c_5 (\tau_3)_{ij} (\rho_p-\rho_n)}{f_\pi^2}
-\frac{1}{f_\pi^2}\frac{\partial {\cal E}_3}{\partial m_\pi^2}~. 
\label{ratio}
\end{align}

\begin{figure}[ht]
\psfrag{qc}{\small $\frac{\la \Omega|\bar{q}q|\Omega \ra}{\la 0|\bar{q}q|0\ra}$}
\psfrag{rho}{\small $\begin{array}{c}\\ \rho~(\hbox{fm}^{-3}) \end{array}$}
\centerline{\epsfig{file=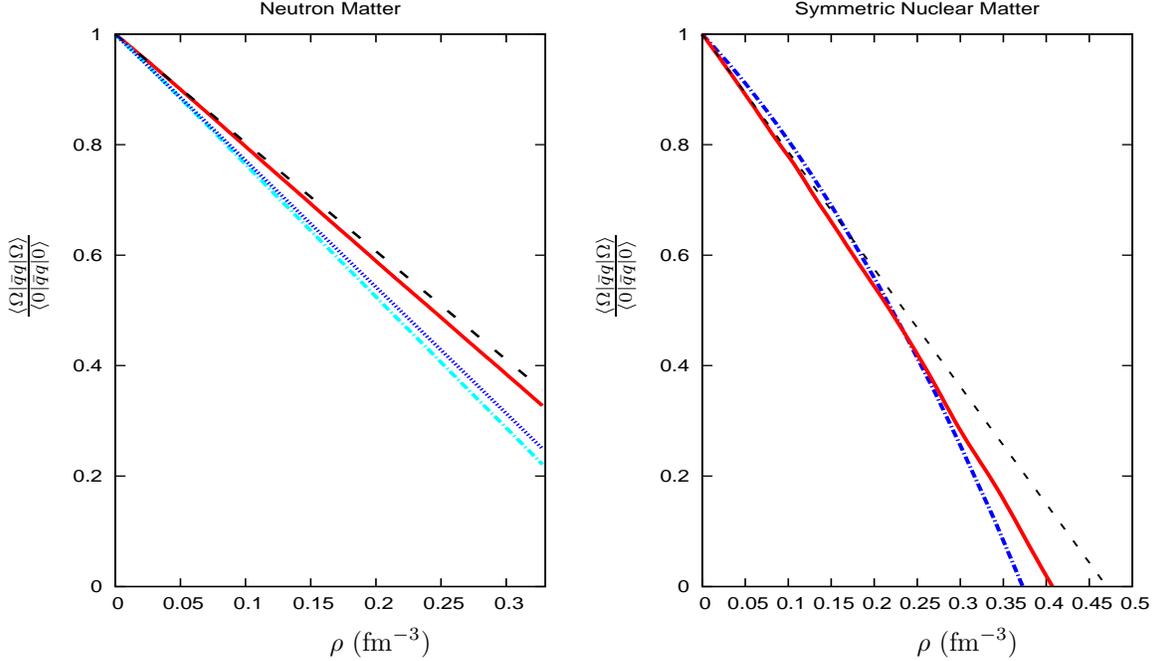,height=.7\textheight,width=.5\textwidth,angle=-90}}
\vspace{0.2cm}
\caption[pilf]{\protect \small
(Color online.) The ratio between the in-medium and vacuum chiral quark
condensate, $\la \Omega|\bar{q}q|\Omega\ra/\la 0|\bar{q}q| 0\ra$, eq.~\eqref{ratio},   
for neutron matter, left panel, and symmetric nuclear matter, right panel. 
Left panel: The (red) solid and (cyan) dot-dashed lines are 
our full results for $\la \Omega|\bar{u}u|\Omega\ra$ and 
$\la \Omega|\bar{d}d|\Omega\ra$, respectively. The (black) dashed and (blue) 
dotted lines correspond in the same order to the linear density approximation ($\Xi_1$) only. Right
Panel: The (red) solid and (black) dashed lines are the full results (with   $g_0=-0.97~m_\pi^2$) and the linear approximation, respectively. 
The (blue) dot-dashed line is the calculation for $g_0=-0.5~m_\pi^2$. All the curves
 shown employ  $\sigma=45$~MeV \cite{Gasser:1990ap}. 
\label{fig:res}}
\end{figure}

The input parameters are  $g_A=1.26$, $f_\pi=92.4$~MeV, $m_\pi=138$~MeV,
$m_N=939$~MeV and $\sigma=45$~MeV. In addition the values of the subtraction constants $g_0$
that reproduce better the equation of state of symmetric nuclear matter 
and neutron matter are used. These values are \cite{Lacour:2009ej} 
$g_0=-0.97$ and $g_0=-0.62~m_\pi^{-2}$, respectively. 
In fig.~\ref{fig:res}  we show the ratio $\la \Omega|\bar{q}q|\Omega\ra/\la 0|\bar{q}q|0\ra$ eq.~\eqref{ratio}
 for pure neutron (left panel) and for symmetric nuclear matter (right panel). 
From the figure  we observe that the leading order correction $\Xi_1$, eq.~\eqref{xi1}, is the dominant contribution. This gives rise to the linear density approximation represented by the 
dashed and dotted lines in the left panel and by the dashed line in the right one.
The NLO corrections rising from $\Xi_6$, eq.~\eqref{xi6.f}, for the case of neutron matter correspond 
to the difference between the solid and dashed lines and between the dot-dashed and dotted lines 
in the left panel of fig.~\ref{fig:res}. The former couple of lines refer to  $\la \Omega|\bar{u}u|\Omega\ra$ and the latter to $\la \Omega|\bar{d}d|\Omega\ra$. In this case the NLO corrections are small and amount to  4.5\% of the LO ones at $\rho=0.3$~fm$^{-3}$. The corrections increase with density, as expected, since higher
three-momenta are available for larger Fermi momentum.  For the case of symmetric nuclear matter the NLO contributions are larger, though still mild. They correspond to the difference between the solid and dashed lines in the right panel 
of fig.~\ref{fig:res} and they are 10\%  
and 20\%  of the leading correction for $\rho=0.3$ and $0.5$~fm$^{-3}$,
respectively. The NLO corrections, $\Xi_6$, tend
to speed up the tendency towards a vanishing  quark condensate in the nuclear
medium (a signal of a possible restoration of chiral symmetry in nuclear
matter). It is worth pointing out that the dependence on the subtraction
constant $g_0$  of the in-medium quark condensate to NLO is just at the level 
of a few per cent. This is shown in 
fig.~\ref{fig:res} by the dot-dashed line where $g_0=-0.5~m_\pi^2$ is used. 
For the reasons discussed, the quark condensate is significantly less dependent on $g_0$ than $E/A$.
 Were $\sigma=64$~MeV  used~\cite{Pavan:2001wz}, the dominant linear density contribution will lead 
to a faster decrease of the in-medium quark condensate. E.g.\ at this level of
approximation, the dashed line in the right panel of fig.~\ref{fig:res} would cross 
the zero at  already $\rho\approx 2\rho_0$.
To this result one should add the difference between the solid and dashed
lines corresponding to the contributions 
from the in-medium nucleon-nucleon contributions, $\Xi_6$, which are
independent of the value of $\sigma$ taken.

The one-pion exchange plus the contact nucleon-nucleon interaction terms 
from ${\cal L}_{NN}^{(0)}$, eq.~\eqref{lnn}, 
are fully iterated, as represented by the ellipsis in the diagrams 4--6 of 
fig.~\ref{fig:all_imcqc}. In ref.~\cite{Kaiser:2007nv} one-pion exchange is iterated
only once, and no nucleon-nucleon contact interactions at the same chiral
order are included. However, the contribution of ref.~\cite{Kaiser:2007nv} from 
once-iterated one-pion exchange  is notoriously larger than ours. Keeping only
this extra contribution, together with the linear density one,
ref.~\cite{Kaiser:2007nv} finds that the quark condensate vanishes already at 
$\rho \simeq 0.24$~fm$^{-3}$, while in our case this happens for the larger
$\rho=0.41$~fm$^{-3}$ (this values coincides very closely to  that for which $E/A$ 
becomes positive in the right panel of fig.~\ref{fig:energy}).  This indicates
that the further iteration of the one-pion exchange, together with the
inclusion of the associated nucleon-nucleon local terms of ${\cal
  L}_{NN}^{(0)}$, have an important impact.  Other mechanisms were included in 
ref.~\cite{Kaiser:2007nv}. They indicate a tendency towards a stabilization of the in-medium quark 
condensate in nuclear matter for densities above $\rho_0$.   However, let us mention that no chiral power counting 
is  followed by the authors in ref.~\cite{Kaiser:2007nv} but an expansion in the number of loops. The latter does not coincide with a chiral power counting expansion because of the infrared enhancement associated 
with the nucleon propagators, as discussed above and also referred to in \cite{peripheral}. In addition, 
the local nucleon-nucleon interactions are not treated consistently within a chiral power counting either. 
Then, we consider as an interesting future task to
include the higher orders needed within our power counting and confirm, if possible, the far reaching
results of ref.~\cite{Kaiser:2007nv}.    Other papers also find that the linear decrease 
with density of the in-medium quark condensate is softened  
for $\rho\gtrsim \rho_0$ \cite{lutz,tubinguen,li}. However, important
differences persist at the quantitative level. E.g.\ ref.~\cite{tubinguen}
finds  
for densities above $\rho_0$ a much milder 
positive correction to the linear density decrease of the quark condensate 
than refs.~\cite{lutz,li,Kaiser:2007nv}. Note that we obtain a realistic
equation of state for symmetric nuclear matter and the quark condensate keeps
its trend to vanish for higher densities, as also observed in
ref.~\cite{tubinguen}.  
We agree on the observation performed in refs.~\cite{Kaiser:2007nv,tubinguen} 
that the short-range nucleon-nucleon interactions effects are  suppressed 
for the evaluation of the in-medium quark condensate which is dominated by
long-ranged pion physics. This is a consequence of eq.~\eqref{derivative.t}. 
At the order we are working, if only the local nucleon-nucleon interactions 
from ${\cal L}_{NN}^{(0)}$ were kept the derivative of $T_{JI}$ with respect
to $m_\pi^2$ would be zero and $\Xi_6\to 0$. 
In the same way, the limit $\vp^2\to \infty$ is independent of the pion mass 
which suppresses the influence of the short-range 
one-pion exchange part.  It follows from the left panel in fig.~\ref{fig:res} 
that for the case of pure neutron matter  the linear tendency of the quark
condensate is only weakly modified 
by the higher order corrections. Similar results are obtained in
ref.~\cite{Kaiser:2008qu} as well. 
 Let us mention in passing that the approach of
the Munich group~\cite{fritsch,fritsch2,Kaiser:2008qu} has difficulties in providing a good 
reproduction of the equation of state for neutron matter while in our approach
\cite{Lacour:2009ej} it emerges in a quite straightforward way at NLO.

\section{Pion decay constant and the Gell-Mann--Oakes--Renner relation}
\label{sec:im_pdc.disc}

Let us now consider the axial-vector current $A_\mu^i=\bar{q}(x)\gamma_\mu\gamma_5(\tau^i/2) q(x)$, with $q(x)$ a two-dimensional vector corresponding to the light quarks fields and $\tau^i$ the Pauli matrices. For notation we refer to ref.~\cite{annp}.

Due to the presence of the nuclear medium one should distinguish between the couplings of the pion to the spatial and temporal  components of the axial-vector current, denoted by $f_s$ and $f_t$, respectively.
The fact that $f_t \not= f_s$ in nuclear matter and some consequences thereof have already been discussed in \cite{Kirchbach:1993ep,Leutwyler:1993gf,Thorsson:1995rj,Wirzba:1995sh,Kirchbach:1996xy,Pisarski:1996mt,wirzba,Pisarski:1996zv,Kim:2001xu,annp}.
The calculations in \cite{annp} indicate a linear decrease of $f_t$ with density, $f_t=f_\pi(1-(0.26\pm 0.04)\rho/\rho_0)$, where $f_\pi=92.4~$MeV is the weak pion decay constant in vacuum and $\rho_0$ is the nuclear matter saturation density.
This result indicates that it makes sense to use chiral Lagrangians in the nuclear medium up to central nuclear densities.
However, the result so far has been obtained by considering pion-nucleon dynamics only, without 
including  nucleon-nucleon interactions.

\begin{figure}[t]
\psfrag{Vr=1}{{\small $V_\rho=1$}}
\psfrag{Vr=2}{{\small $V_\rho=2$}}
\psfrag{Op4}{{\small ${\cal O}(p^4)$}}
\psfrag{Op5}{{\small ${\cal O}(p^5)$}}
\psfrag{PiWFR}{{\small $\pi$-WFR}}
\centerline{\fbox{\epsfig{file=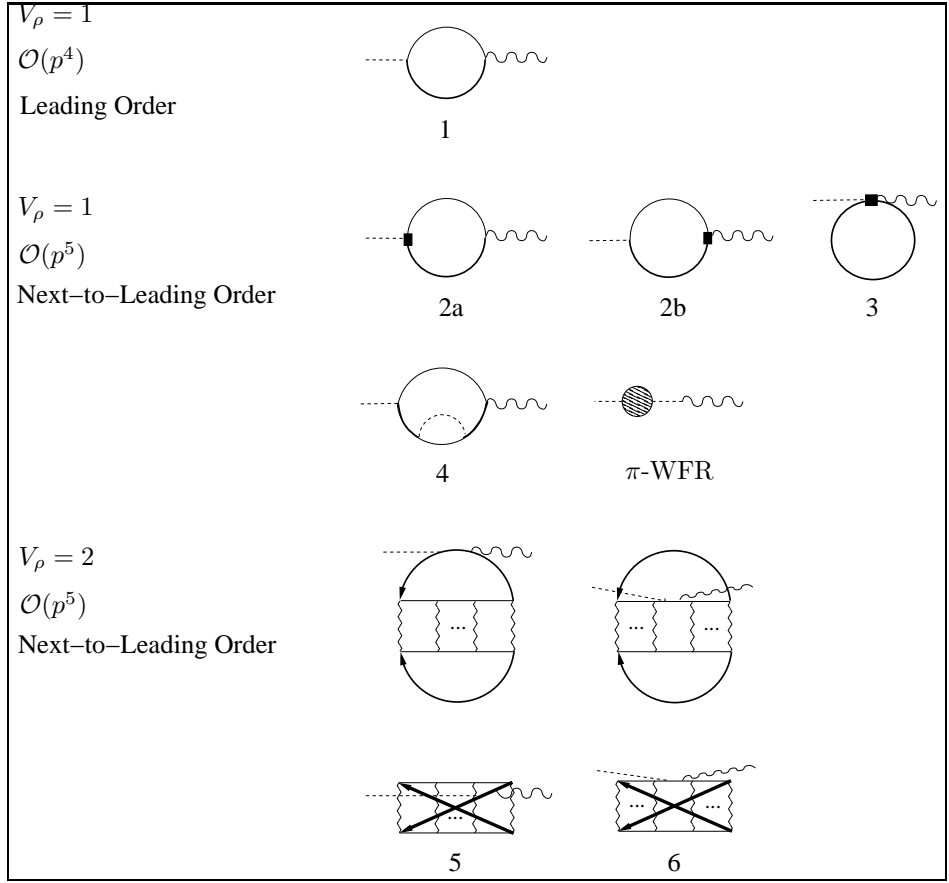,width=.7\textwidth,angle=0}}}
\vspace{0.2cm}
\caption[pilf]{\protect \small
Contributions to the in-medium pion decay up to NLO or ${\cal O}(p^5)$.
The axial vector current is indicated by the wavy line and pions by the dashed ones.
A wiggly line corresponds to the nucleon-nucleon interaction kernel (given in fig.~\ref{fig:wig}) whose iteration is denoted by the ellipsis.
The diagram labelled $\pi$-WFR indicates the contribution from the wave function renormalization of the pion.
Crossed diagrams are not shown.
\label{fig:all_impdc}}
\end{figure} 
We depict the contributions to the process of the in-medium pion decay $\langle\Omega|A_\mu^j|\pi^i\rangle$ in fig.~\ref{fig:all_impdc}.
Diagrams 1--3 were already considered in ref.~\cite{annp}. Regarding diagram~4 one has to take into account the same comments as previously given in section~\ref{sec:im_cqc.pc} concerning diagrams 2 and 3.
 The contribution with only the free part of the nucleon propagators vanishes and that with only 
the density dependent part is taken into account by the diagrams 5 and 6. The remaining contributions, with one density dependent part for one nucleon propagator and a free one for the other, is suppressed by one order.   The reason is because it implies the derivative  of $\Sigma_f^\pi$ with respect to energy. This already occurred for the diagram 2 
of fig.~\ref{fig:all_imcqc} in section~\ref{sec:im_cqc.pc}, see also refs.~\cite{nlou,Lacour:2009ej} for the case of the pion self-energy.  
For the diagrams~5 and 6 we find a mutual cancellation.  In the same way as we  already found that 
for the case of the isovector contributions to the pion self-energy of diagrams~(b) and (d) in
fig.~4 of
\cite{nlou}.
 Because of this  there are no new contributions from the pion wave function renormalization (diagram indicated by $\pi$-WFR in fig.~\ref{fig:all_impdc}) beyond those already considered in ref.~\cite{annp}.  The driving mechanism for such cancellation has been explained here in section~\ref{sub:im_cqc.xi4.5} for the case of the chiral quark condensate involving the mutual cancellation of $\Xi_4$ and $\Xi_5$. 
Due to suppression and cancellation of new contributions, we find that there are no additional contributions to those already given in \cite{annp} eq.~(4.24) for the in-medium pion decay.


Together with the relation for the chiral quark condensate, eq.~\eqref{ratio}, we can write down the Gell-Mann--Oakes--Renner (GMOR) relation for \textit{symmetric} nuclear matter \cite{annp}
\begin{equation}
 \tilde m_\pi^2  f_t^2 = - \hat m \langle\Omega|\bar uu + \bar dd|\Omega\rangle + \delta_0 ~,
\end{equation} 
where  $\tilde m_\pi$ is the in-medium pion mass and $\delta_0$ corresponds to corrections that  start at ${\cal O}(p^4)$ in vacuum $\chi$PT, which implies that the GMOR relation is only exact at lowest order.
The stability of the GMOR relation under the in-medium corrections as well as the fact that it is the temporal coupling $f_t$, and not the spatial one $f_s$, the one involved in the GMOR relation has previously been reported in \cite{Thorsson:1995rj,Wirzba:1995sh,Kirchbach:1996xy,wirzba,annp} within the mean field approximation, and in \cite{Kim:2001xu} in the framework of QCD sum rules.
Expanding the in-medium contributions to the pion mass, the pion decay constant and the chiral quark condensate explicitly we find
\begin{equation}
 m_\pi^2 f_\pi^2 \, \Bigl( 1 + \delta_{m_\pi^2}^{(2)} + \delta_{m_\pi^2}^{(3)} + \ldots \Bigr) \Bigl( 1 + 2\delta_{f_\pi}^{(2)} + 2\delta_{f_\pi}^{(3)} + \ldots \Bigr) = - \hat m \langle 0|\bar uu + \bar dd|0\rangle \, \Bigl( 1 + \delta_{\Xi}^{(3)} + \delta_{\Xi}^{(4)} + \ldots \Bigr) + \delta_0 ~,
 \label{eq:im.gmor}
\end{equation}
where  $\delta_{m_\pi^2}^{(i)}$, $\delta_{f_\pi}^{(i)}$ and $\delta_{\Xi}^{(i)}$ denote the in-medium corrections of the pion mass squared, pion decay constant and chiral quark condensate, respectively, and the superscript indicates the corresponding chiral order.
The relative $\cO(p^2)$ corrections on the left-hand-side of eq.~\eqref{eq:im.gmor}, $\delta_{m_\pi^2}^{(2)}$ and $\delta_{f_\pi}^{(2)}$, only appear for the charged pions and correspond to isospin breaking due to different proton and neutron densities in asymmetric nuclear matter \cite{annp}.
These in-medium corrections to the pion mass and the pion decay constant vanish for symmetric nuclear matter.
They also cancel if we take the average of the pion masses or accordingly the pion decay constant.
 For these reasons, we can neglect the relative corrections at $\cO(p^2)$ for the consideration of the GMOR relation.
As previously discussed, up to next-to-leading order the leading nucleon-nucleon contributions to the pion decay, and therefore to the decay constant, cancel in $\delta_{f_\pi}^{(3)}$.
The same observation was already made for the pion self-energy in \cite{nlou} and, therefore, contributions due to nucleon-nucleon interactions are absent in $\delta_{m_\pi^2}^{(3)}$ as well.   Regarding the right-hand-side of eq.~\eqref{eq:im.gmor}, $\delta_{\Xi}^{(3)}$ has been calculated in section~\ref{sec:im_cqc.pc} and corresponds to $\Xi_1$, eq.~\eqref{xi1}. In contrast to $\tilde m_\pi^2$ and $f_t$ there is 
a non-vanishing NLO in-medium correction to the quark condensate due to the in-medium nucleon-nucleon 
interactions and given by  
$\Xi_6$, eq.~\eqref{xi.c.2}. Thus,  $\delta_\Xi^{(4)}\neq 0$. However, this is  
${\cal O}(p^6)$, one order above the contributions discussed for the 
left-hand-side of eq.~\eqref{eq:im.gmor}. Whence, it would be needed a full N$^2$LO calculation for $f_t$ and $\tilde m_\pi^2$ in order to 
ascertain the stability of the in-medium corrections to the GMOR relation up to ${\cal O}(p^6)$, which is beyond the scope of the present work. 
On the other hand, as shown in \cite{annp}, the pion-nucleon dynamics  
 that gives rise to 
$\delta_{m_\pi^2}^{(3)}$, $\delta_{f_\pi}^{(3)}$ and $\delta_{\Xi}^{(3)}$ for symmetric nuclear matter does  not violate the in-medium GMOR relation, eq.~\eqref{eq:im.gmor}.
Then, we conclude that  the in-medium corrections, including nucleon-nucleon interactions, do not spoil the validity of the Gell-Mann--Oakes--Renner (GMOR) relation up-to-and-including ${\cal O}(p^5)$ (or NLO) in our 
in-medium power counting scheme. 

\section{Conclusions and outlook}
\label{sec:im_cqc.conc}

Employing the  in-medium power counting of \cite{nlou} and the methods of Unitary $\chi$PT for taking into account the resummation of non-perturbative effects \cite{Lacour:2009ej} we have calculated  the chiral quark condensate up-to-and-including next-to-leading order, ${\cO}(p^6)$ in nuclear matter.
We have found an interesting partial cancellation between the diagrams
involving the nucleon-nucleon interactions. In this way, those contributions
that arise due to the leading quark mass dependence of the nucleon mass
mutually cancel. This corresponds to the diagrams 4 and 5 in
fig.~\ref{fig:all_imcqc}. As a result, only the diagrams 6 in the figure, that stem
from the quark mass dependence of the pion mass, survive. This is the reason
why previous calculations have found that short range nucleon-nucleon
interactions are suppressed for the calculation of the in-medium quark
condensates,  which is then dominated by the long-range pion contributions. 
Note that we did not exploit the Hellmann-Feynman theorem, but obtained our results explicitly from the generating functional, eq.~\eqref{g.qc}.
We conclude that the corrections are small for the quark condensate in the case of pure neutron matter. These corrections are more significant for the the case of symmetric nuclear matter. It is also worth pointing out that the full iteration of the nucleon-nucleon interactions reduces the force of such extra damping of the quark condensate as compared with other references. The dependence of our results on the subtraction constant $g_0$ is at the level of a few percent, much smaller than for the case of the binding energy \cite{Lacour:2009ej}. On top of these 
nuclear effects a reduction of the present uncertainty in the pion-nucleon sigma-term would be most welcome.

We have also addressed the calculations of the pion decay constant and tested the stability of the GMOR relation against in-medium  NLO in our chiral counting, that is, up to ${\cal O}(p^5)$.
The nucleon-nucleon contributions for the in-medium corrections vanish at this order, not only for the calculation of the pion mass but also for the in-medium pion decay constant.
This implies that the expressions in ref.~\cite{annp} for the coupling of the pion to the axial-vector 
current are right up to NLO, despite having considered the inclusion of nucleon-nucleon corrections. 
 Because of this result we also find that in-medium contributions do not spoil the GMOR relation 
up to ${\cal O}(p^5)$, as in ref.~\cite{annp}. The worked out NLO contributions to the quark condensate 
give rise to a new contribution to the GMOR relation at ${\cal O}(p^6)$ due to the nucleon-nucleon interactions that would require a N$^2$LO calculation for the pion mass and weak pion decay constant in symmetric nuclear matter in order to check the stability of the GMOR relation. This would be an interesting future task.

Higher order calculations are of most interest  as they would provide the important two-pion exchange and multi-nucleon forces.
 They would  furthermore merge meson-baryon mechanisms with novel multi-nucleon contributions that can be worked out systematically within our EFT.
The large  impact of the $\Delta$-isobar excitation to symmetric nuclear matter in \cite{Kaiser:2007nv}, which is then partially compensated by the three-body interactions terms proportional to the low-energy constant $c_1$, freezes the dropping of the in-medium chiral quark condensate at $\sim 2\rho_0$. This interesting fact  requires confirmation within our power counting so as to keep all the terms contributing at the same chiral order, while keeping the full iteration of lowest order local and one-pion exchange diagrams. As commented above, the latter has a significant impact in the contribution at NLO. 
In addition, one should keep in mind that the proper way to address the issue of chiral symmetry restoration in the nuclear medium is the calculation of the  temporal pion decay constant in the nuclear medium \cite{annp}, which should be pursued for higher orders.


\section*{Acknowledgements}
AL wants to thank Andreas Wirzba for useful discussions.
This work is partially funded by grants MEC  FPA2007-6277, Fundaci\'on S\'eneca grant 11871/PI/09 by 
BMBF grants 06BN411 and 06BN9006, EU-Research Infrastructure
Integrating Activity
 ``Study of Strongly Interacting Matter" (HadronPhysics2, grant n. 227431)
under the Seventh Framework Program of EU 
and HGF grant VH-VI-231 (Virtual Institute ``Spin and strong QCD'').


\end{document}